\documentclass[aps,reprint,amsmath,amssymb]{revtex4-2}
\usepackage{graphics}
\usepackage{subfigure}
\usepackage{epsfig}
\usepackage{epsf,epic}
\usepackage{dcolumn}
\usepackage{color}
\usepackage{marvosym}
\usepackage{subfigure}
\usepackage{amsmath}
\usepackage{amssymb}
\usepackage{amsfonts}
\usepackage{wrapfig}
\usepackage{pstricks}
\usepackage{multirow}
\usepackage{bm}
\graphicspath{{./figs/}}
\usepackage{dcolumn}
\newcommand{\etal}{\textit{et al.\ }}

\begin{document}
\title{Computational study of defect complexes in  $\beta$-LiGaO$_2$ and  their relation to the donor-acceptor-pair recombination.}

\author{Klichchupong Dabsamut}
\author{Adisak Boonchun}
\affiliation{Department of Physics, Faculty of Science, Kasetsart University, Bangkok 10900, Thailand}
\author{Walter R. L. Lambrecht}\email{walter.lambrecht@case.edu}
\affiliation{Department of Physics, Case Western Reserve University, 10900 Euclid Avenue, Cleveland, Ohio 44106-7079, USA}
\begin{abstract}
  Hybrid functional calculations are presented for defects in LiGaO$_2$
  with the fraction of non-local exchange adjusted to reproduce
  the recently reported exciton  gap of 6.0 eV.  We study how the
  defect transition levels of the main native defects change
  with respect to the band edges compared to earlier calculations which
  assumed a smaller band gap near 5.1 eV.  In addition, we consider
  defect complexes formed by combining the main native donor Ga$_\mathrm{Li}$
  with the main acceptors, $V_\mathrm{Li}$ and Li$_\mathrm{Ga}$ antisites
  as function of their relative position.  These results are used
  to tentatively identify the photoluminescence bands
  previous assigned to donor-acceptor-pair recombination. 

\end{abstract}
\maketitle
\section{Introduction}
Lithium gallate ($\beta$-LiGaO$_2$)  
has recently been proposed as a potential ultra-wide-band-gap (UWBG) 
semiconductor. Its crystal structure, reported  by Marezio \cite{Marezio65},
is a 
cation-ordered wurtzite-derived structure with space group $Pna2_1$. Large heteroepitaxial
bulk single crystals, grown by the 
Czochralski method have been reported \cite{Ishii98,Chen14}
and most of this effort was motivated by the need for lattice matched
substrates for heteroepitaxial growth of GaN \cite{Ishii98,Christensen05,Doolittle98}. Its thermal properties are of interest in this context
and were reported by by Weise and Neumann \cite{Weise96} and  Neumann \etal\cite{Neumann87}. It can also be used as substrate for 
heteroepitaxial growth of  ZnO\cite{Ohkubo2002}.
In fact, ZnO can be considered as the wurtzite parent compound of
LiGaO$_2$, which conceptually is obtained by replacing the group-II Zn ion
by alternating group-I Li  and group-III Ga ions, thereby locally
maintaining the octet-rule if each O is surrounded by two Li and two Ga ions.
It can thus be alloyed with ZnO  \cite{Omata08,Omata11}
and other ternary I-III-O$_2$ oxides
like CuGaO$_2$ \cite{Suzuki2019}, which opens the way toward band gap tuning. 

Until recently LiGaO$_2$ was primarily considered as an optical material
or substrate material for growth of other semiconductors. 
Its elastic, phonon and piezoelectric properties were calculated
using density functional theory (DFT) by Boonchun and Lambrecht \cite{Boonchun10}. It electronic structure was calculated
 at the DFT level using  the modified Becke-Johnson (mBJ) exchange-correlation \cite{Becke06,TranBlaha09} 
functional by Johnson \etal\cite{Johnson2011}.  Its optical gap was obtained from absorption measurements \cite{Wolan98,Chen14}
and a combination of X-ray absorption and emission spectroscopies \cite{Johnson2011} and found generally to be about 5.3--5.6 eV. Recently,
quasiparticle self-consistent (QS)$GW$ (where $G$ is the one-electron Green's function and $W$ the screened Coulomb interaction) calculations were
performed by Radha \etal \cite{Radha21} for various
crystal structures of LiGaO$_2$ following an earlier
QS$GW$ calculation by Boonchun \etal \cite{Boonchun11SPIE}.

The possibility of
doping the material and thereby making a functional semiconductor was first
suggested by Boonchun and Lambrecht\cite{Boonchun11SPIE} but in this
study only small cells with unrealistically high dopant concentrations
were considered. 
Its native defects were recently studied using first-principles
calculations \cite{Boonchun19}. The  opportunities for
$n$-type and $p$-type doping were  studied in Ref. \cite{Dabsamut20}.
It was predicted that Si and Ge would be shallow donors, while Sn would be a deep donor.  However, $p$-type doping by N on O or by Zn on Ga
site were found to be less promising because of deep levels and site competition
with Zn on Li donors. Doping by various diatomic molecules was also investigated
but not found to lead to $p$-type doping \cite{Dabsamut22}. Electron paramagnetic resonance of Li and Ga vacancies
were reported by Lenyk \etal \cite{Lenyk18} and analyzed computationally by Skachkov \etal \cite{Skachkov20}.

Recently, the infrared (phonon related) as well  as visible ultraviolet (interband transition related)
optical properties were studied by  reflectivity, transmission and spectroscopic reflectivity by Tum\.enas \etal\cite{Tumenas17}
and indicated sharp excitons near 6.0 eV. Luminescence properties were studied by Trinkler \etal\cite{Trinkler17,Trinkler22}
and the photoluminescence excitation (PLE) spectroscopy confirmed the presence of sharp free excitons near 6.0 eV.
The anisotropic splitting of these excitons, reported in \cite{Trinkler22} reflects the valence band splitting, characteristic
of the orthorhombic symmetry of the crystal and is in good agreement with the recent computational study
by Radha \etal \cite{Radha21}. This much larger optical exciton gap
than previously accepted led one of us to re-examine the convergence
of the QS$GW$ calculations and to also study the excitons by means of the
Bethe-Salpeter-Equation method and not only found resuls in close agreement
for the exciton gap near 6.0 eV but also found a large exciton binding
energy of about 0.7 eV. \cite{Dadkhah23} Furthermore, a series of excited state excitons were revealed in this work.
In view of this larger gap we  here present new calculations of some
of the primary defects found earlier but with the hybrid functional
fraction of exchange adjusted to the larger gap.  Furthermore
the work of Trinkler et al. \cite{Trinkler17}
assigned the primary photoluminescence
peaks to donor acceptor pair recombination.  This inspired us to
consider defect complexes combining donors and acceptors.

 Trinkler \etal\cite{Trinkler17} assigned
a peak in luminescence centered at about 4.43 eV and from which the PLE gave sharp peaks near 6.0 eV to a donor acceptor pair (DAP) 
recombination  based on its blue shift under higher power excitation, and temporal behavior. Our prior study \cite{Boonchun19} of defects identified
the Ga$_\mathrm{Li}$ as the primary native defect donor and $V_\mathrm{Li}$ and Li$_\mathrm{Ga}$ as the most likely acceptors with the
second one having slightly higher acceptor binding energy. However, those calculations were done with a hybrid functional scheme
in which the non-local exchange fraction $\alpha$ was set to 0.25 which gave a band gap of 5.1 eV. In view of the now established significantly larger gap, we repeated these calculations for a larger fraction of exchange ($\alpha=0.38$) tuning the gap to 6.09 eV and also considered
close pairs of donors and acceptors in a complex.

\section{Computational Method}
Our calculations were performed using the VASP package\cite{VASP1,VASP2}  using the projector-augmented wave (PAW) method\cite{PAW} and the Heid-Scuseria-Ernzerhof (HSE) hybrid exchange correlation functional \cite{HSE03,HSE06} with screening length of 10 \AA\  and fraction of non-local exchange chosen $\alpha=0.38$ chosen to adjust the band gap. 
Other aspects of our computational method are the same as in Refs. \cite{Boonchun19,Dabsamut20,Dabsamut22}.
While the 6.09 eV corresponds strictly to the exciton gap and not to the quasiparticle gap
one may view these calculations as a guide for how the defect levels behave with the gap and assume that excitonic effects are
pertaining to the defect levels as well. In other words, donor or acceptor bound excitons are assumed to have similar exciton binding
energy as the free exciton.
\section{Results and Discussion}

\begin{figure}
  \includegraphics[width=9cm]{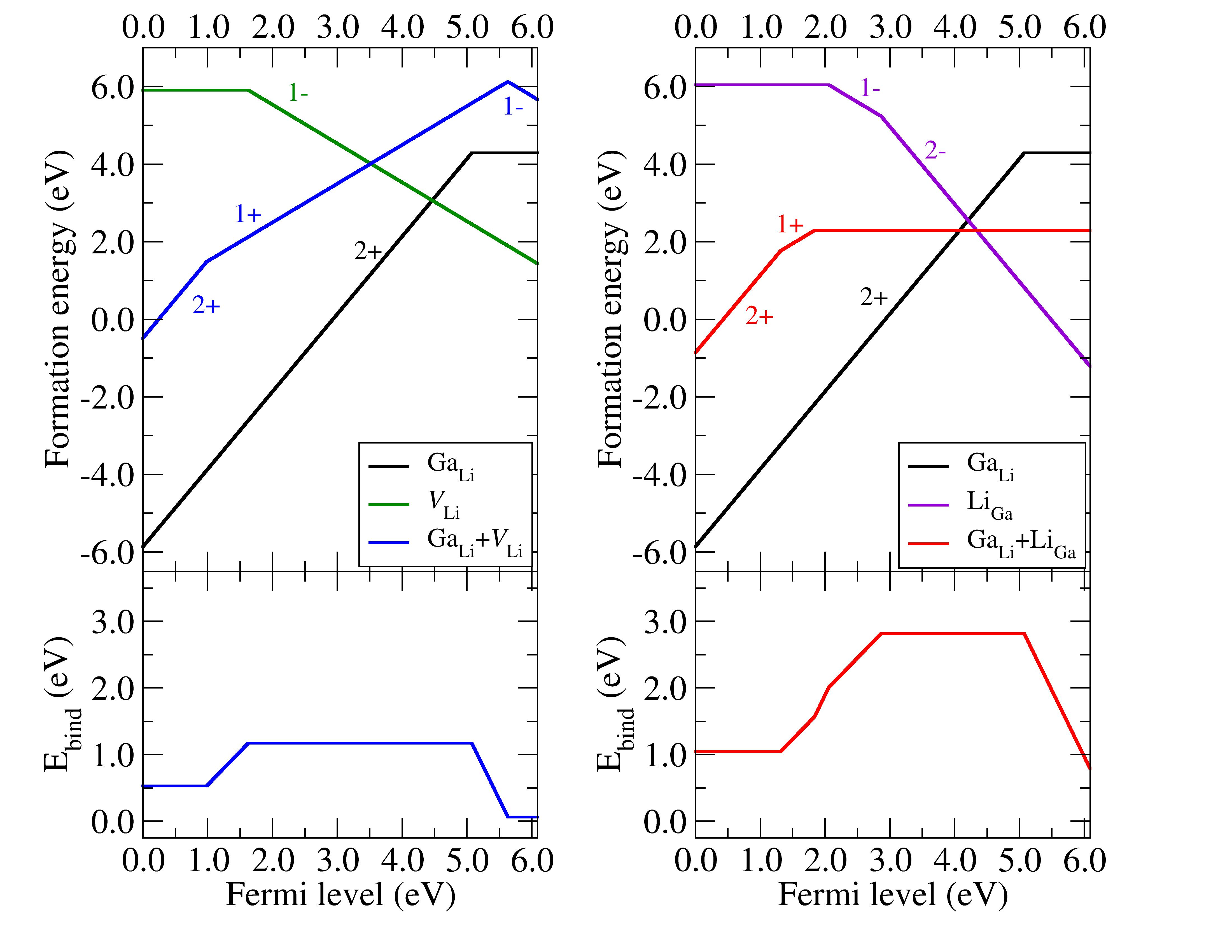}
  \caption{Top panel : Donor and acceptor transition levels and their nearest neighbor complexes. Bottom panel : Complex defects binding energy. Left: $V_\mathrm{Li}$ and right Li$_\mathrm{Ga}$ acceptor combined with the same donor Ga$_\mathrm{Li}$}
  \label{figdefect} 
\end{figure}

\begin{table}[h]
	\caption{
		Defect transition levels $\varepsilon(q,q')$ in eV with respect to the VBM.}
	\label{tablevels}
        \begin{ruledtabular}
	\begin{tabular*}{0.4\textwidth}{@{\extracolsep{\fill}}ccc} 
		Defect & $q,q'$ & $\varepsilon(q,q')$  \\ \hline
		Ga$_\mathrm{Li}$  & (2+/1+) & 5.26 \\
		& (1+/0) & 4.89 \\
		& (2+/0) & 5.08 \\
		$V_\mathrm{Li}$  & (0/1-) & 1.63 \\
		Li$_\mathrm{Ga}$ & (0/1-) & 2.06 \\
		& (1-/2-) & 2.87 \\
		Ga$_\mathrm{Li}$+$V_\mathrm{Li}$ & (2+/1+) & 0.98 \\
		& (1+/1-) & 5.63 \\
		Ga$_\mathrm{Li}$+Li$_\mathrm{Ga}$ & (2+/1+) & 1.32 \\
		& (1+/0) & 1.84 \\
	\end{tabular*}
        \end{ruledtabular}
\end{table}

The results for the main defects of interest here are shown in Fig. \ref{figdefect} and the transition levels are summarized in
Table \ref{tablevels}.
It is important to note that compared to our previous study of native
defects \cite{Boonchun19} both donor and acceptor levels moved deeper into
the gap when using a larger fraction of exchange and hence larger gap.
In that study, the gap was 5.1 eV and the V$_\mathrm{Li}$ $(0/-)$ level was
located at 1.03 eV above the VBM, the Li$_\mathrm{Ga}$ $(0/-)$ acceptor at 1.55 eV
above the VBM and the Ga$_\mathrm{Li}$ $(2+/0)$ deep donor level at 0.74 eV below
the CBM. However, the difference between donor and acceptor levels
were close to what we find here.

As in our previous work we find the Ga$_\mathrm{Li}$ to be a deep double donor with a $2+/0$ transition. This implies a so-called negative $U$ center
in which the $q=+1$ state has higher energy than the neutral or $2+$ state for any Fermi level position. However, in photoluminescence, we are not dealing with equilibrium but with a situation where electrons are excited to the conduction band and then relax to become trapped on the donors, which initially were in the $q=+2$ state. Thus some donors will occur in the $q=+1$ state and may then recombine with holes before they trap a second electron and become neutral. 
It is therefore also important to locate the $2+/+$  level. We find the $2+/0$ to lie at 1.01 eV below the CBM and the $2+/1+$ at 5.26 eV (above VBM) or 0.83 eV below the CBM 
and the $1+/0$ at  VBM+4.89  or CBM$-$1.2 eV. Meanwhile the $V_\mathrm{Li}$ vacancy $0/-$ level lies at about 1.63 eV above the VBM.  The energy difference between these
remote and isolated donor and acceptors is thus 3.45 eV or 3.63 eV depending on whether we use the $2+/0$ or $2+/+$ level of the Ga$_\mathrm{Li}$.
On the other hand, in a DAP transition the photon energy is given by
\begin{equation}
  \hbar\omega=E_g-E_D-E_A+e^2/\varepsilon R,
\end{equation}
where $R$ is the distance between the
donor and acceptor and $\varepsilon$ the dielectric constant. This is because in the final state after recombination, the donor and acceptor find themselves in ionized states and experience a Coulomb attraction. This energy gain is transferred to the photon. Usually, for shallow donors and acceptors one  uses the static dielectric constant which includes lattice screening.  Here, however the donor and acceptor are both quite deep
and hence the binding energies may involve only electronic screening. In other words, the electron and hole involved in the recombination move too fast around their donor and acceptor for the lattice vibrations to participate in the screening of their Coulomb interaction. A full calculation of the DAP spectrum is rather complex as it would involve the random probability distribtion of the
donors and acceptors relative to each other and the overlap of their wave functions for each separation, which would determine their
recombination probability.  Instead we start from the experimental value of the peak position and estimate which distance this
would correspond to. 
To explain the DAP peak at 4.4 eV we need to assume the $e^2/\varepsilon R\approx0.8$
eV if we assume the donors make a transition from $+$ to $2+$.
 Using a  isotropically averaged dielectric high-frequency constant \cite{Tumenas17,Dadkhah23} of 3.0, this suggests
a distance $R\approx6.0$ \AA\  between the donor and acceptor corresponding
to the peak value of the DAP band. 
 The spectrum of this DAP is asymmetrically stretched with a tail toward
lower recombination energies which would correspond to more distant pairs.
This DAP distance of 6.0 \AA\ is slightly larger than third nearest neighbor $V_\mathrm{Li}-\mathrm{Ga}_\mathrm{Li}$
pairs. The closest distance between them is 3.10 \AA. Trinkler \etal\cite{Trinkler17} proposed a randomly placed DAP  for the 4.43 eV band and assign it to
tunnel recombination.  Keeping in mind that with such rather deep acceptor and donors the overlap of their wave functions which would allow tunneling
would be rather small for remote DAP.  A distance of $\sim$ 6 \AA\  is thus not unreasonable for the peak position.

We can see that the Li$_\mathrm{Ga}$  acceptor level is 0.44 eV deeper than the $V_\mathrm{Li}$.  Hence with similar assumptions, this DAP
would then occur at about 3.9 eV if we assume a similar distance between  donor an acceptor corresponding to the peak of the DAP, as for the $V_\mathrm{Li}$.  Hence these acceptors could possibly account for the second observed luminescence band peaked at 3.76 eV.
Trinkler \etal \cite{Trinkler17} also suggested that the 4.43 eV could in part also result from a fast decay related to a free electron to acceptor
(eA) process. With our calculated acceptor levels of $V_\mathrm{Li}$ at 1.63 and Li$_\mathrm{Ga}$ at 2.06 eV
 above the VBM and the gap of 6.09 eV, this would correspond to 4.46 and 4.03 eV, which are indeed also close to the
observed PL band peak positions.

We here also explicitly consider donor acceptor complexes at various
relative positions from each other.
The bottom part of Fig. \ref{figdefect} shows that there is a net binding
energy between the donor and acceptor in the DAP complex over a considerable
range of Fermi energies which is defined by Equation 2 in \cite{defectinsolid}, 
or $E_B=-E_f(DAP)+E_f(D)+E_f(A)$ where the DAP complex and the donor D and acceptor A formation energies are all 
taken at the same Fermi-level position. 

\begin{figure}
  \includegraphics[width=9cm]{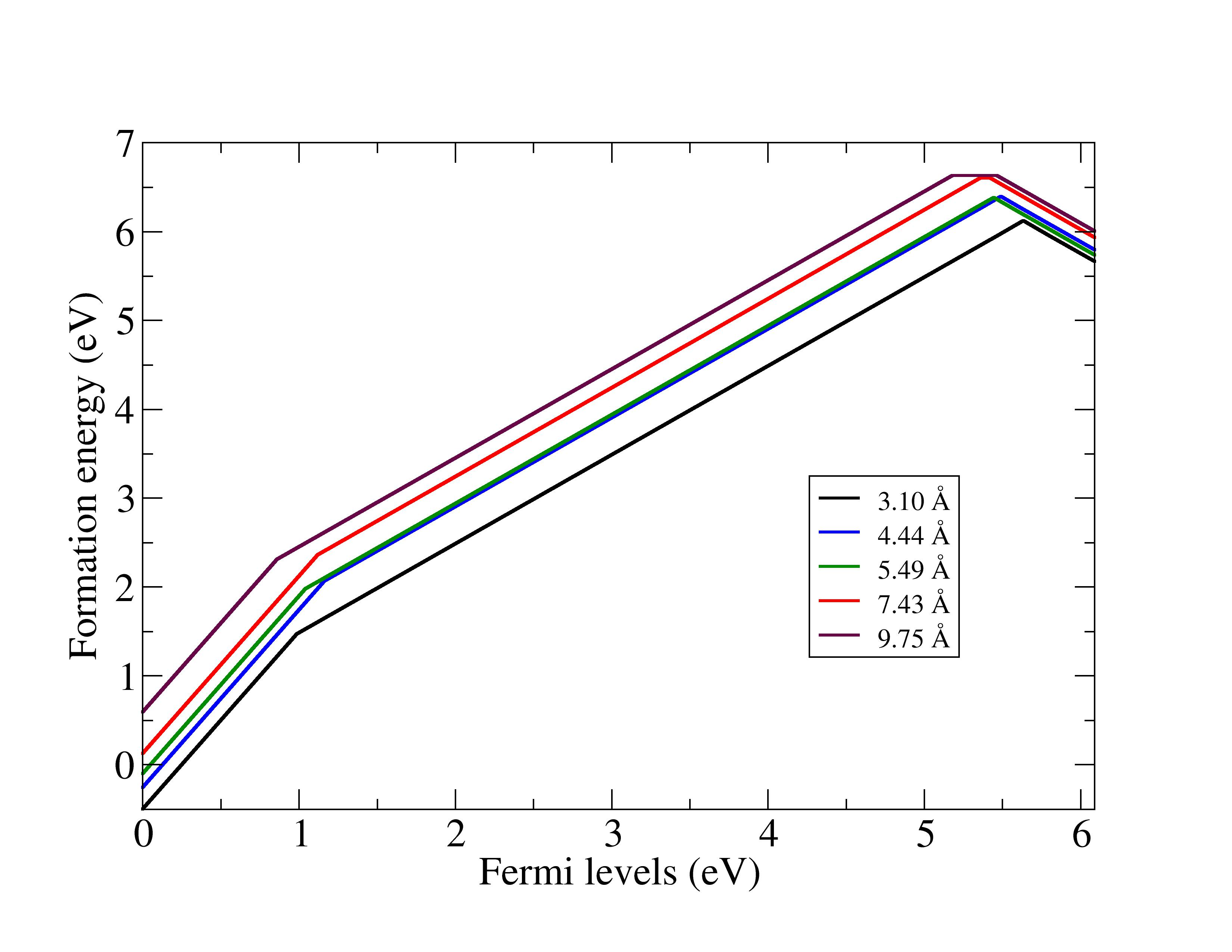}
  \caption{Transition levels of $V_\mathrm{Li}-\mathrm{Ga}_\mathrm{Li}$
    complexes for various distances between the donor and acceptor parts
    of the complex.\label{vligalid}}
\end{figure}

Considering the defect levels of the
nearest neighbor complex in Fig. \ref{figdefect},
we first note that its transition levels are closely related to the
individual donor and acceptor ones. When the acceptor is in the neutral state and the donor in the $2+$ state, for Fermi level positions close to the VBM,
the complex is obviously in a $2+$ state. Now, first the acceptor goes to the $q=-1$ state and the complex than goes to a $+1$ state.  Since the
$V_\mathrm{Li}$ only occurs in 0 and $-1$ states, the next transition happens for the donor part of the complex Ga$_\mathrm{Li}$ transitioning from 2+ to 0.
At that point the complex will thus go from $+1$ to $-1$. However, we see from Fig.\ref{figdefect} that the $2+/+$ and $+/-$ transitions of the complex
are pushed closer to their respective band edges.  This must result from the defect levels in the complex interacting with each other.  Forming bonding
and antibonding states between the donor and acceptor  wave functions as in a molecule will push these states farther apart.
This means that in a complex the $E_D$ and $E_A$ binding energies  of the neutral DAP before its recombination are reduced, but
this reduction will decrease the farther the D and A are apart. The net result is that $E_D-E_A$ is increased.
For the closest distance of D and A the donor and acceptor energies are about 0.5 eV below the CBM and 
1 eV above the VBM giving a DAP energy of 4.5 eV in the initial state but the Coulomb energy is then about 1.3 eV so the DAP photon energy would be about 
5.8 eV. This is above the maximum of the DAP band, which is near 5 eV, indicating that such close DAPs are unlikely. On the other hand, transitions with energies 
larger that the peak energy up to 5 eV could be partially accounted for by slightly more remote DAPs. In fact for the DAP at 9.75 \AA\ apart (shown in Fig. \ref{vligalid}) the DAP photon energy 
using the D and A transition levels from Fig. \ref{vligalid} and a Coulomb energy at that distance we obtain 4.7 eV, which is still within this DAP band. 

In Fig. \ref{vligalid} we show the transition levels of various
$V_\mathrm{Li}-\mathrm{Ga}_\mathrm{Li}$ complexes with different distances
between the donor and acceptor.  We can see that for the larger
distances, we can now separately see the +/0 and 0/- transitions of the donor
part. However, it is also clear that  the larger the distance the larger
the donor binding energies become.  On the acceptor side, the same trend is
not as clear although still present. This indicates that the interaction
between donor and acceptor levels in the complex does not only depend
on their relative distance but also on their relative orientation. In
other words, anisotropy of the acceptor-like wave function plays a role. 

\section{Conclusions}
In summary,
for the DAP recombination band peaked at 4.43 eV observed in
\cite{Trinkler17},  we can tentatively assign
the donor  to be the native defect Ga$_\mathrm{Li}$ donor
and the $V_\mathrm{Li}$ as the acceptor. The same donor but the Li$_\mathrm {Ga}$ acceptor  could then explain the slightly lower energy 3.76 eV
DAP PL bands.  As already suggested by Trinkler \etal\cite{Trinkler17}
these PL bands would indeed also be consistent with having a contribution from free-electron to acceptor recombinations. Furthermore, we find
that for relatively close DAP distances, there is a considerable repulsion
of the neutral donor and acceptor levels  moving them closer to the band edges
in the initial excited state before the recombination. 

\acknowledgements{This work made use of the High Performance Computing Resource in the Core Facility for Advanced Research Computing at Case Western Reserve University. WRL was supported by the U.S. Air Force Office
  of Scientific Research (AFOSR) under grant no. FA9550-22-1-0201. }

\section*{Conflict of Interest}
The authors have no conflicts to disclose.

 \bibliography{DAP-ligao2.bib}
\end{document}